\documentstyle[aps,prl]{revtex}

\title{Are all noisy quantum states obtained from pure ones?}

\author{Leah Henderson$^{a}$, Noah Linden$^{a}$ and
Sandu Popescu$^{b,c}$}

\address{
$^a$Department of Mathematics, University of Bristol, University
Walk, Bristol, BS8 1TW, UK,\\ $^b$HH Wills Physics Laboratory,
University of Bristol, Tyndall Avenue, Bristol, BS8 1TL, UK,\\
$^c$BRIMS, Hewlett-Packard Laboratories, Stoke Gifford,
Bristol BS12 6QZ, UK.}

\date{11th April 2001}

\begin{document}

\twocolumn

\draft

\maketitle

\begin{abstract}
We ask what type of mixed quantum states can arise when a number
of separated parties start by sharing a pure quantum state and
then this pure state becomes contaminated by noise.  We show that
not all mixed states arise in this way.  This is even the case if
the separated parties actively try to degrade their initial pure
state by arbitrary local actions and  classical communication.

\end{abstract}


Density matrices and the question of their entanglement have been
studied very intensively during the last few years\cite{many}.
However the question of how these density matrices arise in the
first place has received much less attention. It has been tacitly
assumed that density matrices arise when a number of parties,
separated in space, start by sharing a pure state, and then this
state gets contaminated by noise, due to interaction with the
environment. Is it the case that any density matrix can be
obtained in this way? Surprisingly we will show that the answer is
no.

The point is that, in the above scenario, the very fact that we can talk
about separated parties, means that the noise is {\em local}. This imposes
constraints as to how entangled states may be degraded by the environment.
We study here the effects of these constraints.

We will consider two different situations. The first is where we
consider a pure state shared by the parties which is
 contaminated by local noise; this is a typical
situation of obvious physical significance. We will refer to
states which can be produced from pure states by local
contamination as LC states. The second situation is one in which
the parties actively try to degrade the pure state: in addition to
local noise we allow local measurements and classical communication
between the parties. We refer to states which can be produced from
pure states by local contamination and classical communication as
LCCC states.

We note that in both situations, when we wish to get a density matrix
of a given number of parties each with a given dimension of local Hilbert space,
we demand that the precursor pure states  lie in a system Hilbert space of
the same local dimensions\cite{footnote}.

The specific question we address is:
given a general \lq\lq target\rq\rq\ density matrix, can we find a
pure state from which this density matrix can be obtained by local
contamination, in either situation. We will show that generically,
in both situations, the answer is no.

In a sense, density matrices which can be obtained by local
contamination are simple in that their entanglement
properties are simply  related to those of their
pure state precursor. The density matrices  which cannot
be obtained in this way have a more subtle and complex structure.

Let us first discuss the situation in which no classical
communication is allowed. We consider $n$ separated parties each
of which has a $d$ level system. The number of real parameters
describing pure states is $2 d^n - 2$, and the number describing
general mixed states is $d^{2n} -1 $. The fact that it is not
possible to reach an arbitrary \lq\lq target\rq\rq\ density matrix
by degrading any pure state follows simply from the fact that the
number of parameters describing local degrading is {\em linear} in
$n$. Thus for sufficiently large $n$, the number of parameters
describing density matrices will be larger that the number
describing pure states and  local
contamination.

In order to see how large $n$ needs to be we need to calculate the number of
parameters describing the set of general local transformations. The most
general interaction of a $d$-level system with its environment is as
follows \cite{Schumacher}:
\begin{eqnarray}
| i\rangle_S | 0\rangle_E \mapsto |i^\prime\rangle_{SE} = \sum_j |
j\rangle_S | e_{ij}\rangle_E\label{general_local}
\end{eqnarray}
where $| i\rangle_S$ are an orthonormal basis for the states of
the system and $| e_{ij}\rangle_E$ are $d^2$ arbitrary (i.e. not necessarily normalised
or orthogonal) states of the
environment. It is only the norms and overlaps of the $|
e_{ij}\rangle_E$ which are important and there are $d^4$ real
parameters describing these.  In fact the number of parameters
describing the local degradation is less than $d^4$ since the
$d^2$ environment states satisfy $d^2$ conditions arising from the
fact the original states of the system $| i\rangle_S$ are an orthonormal
basis.

Thus the total number of real parameters describing all pure states plus the
number describing the local contamination is at most
\begin{eqnarray}
2 d^n -2 + n (d^4-d^2).\label{LC_parameters}
\end{eqnarray}
This number is to be compared to the number of parameters describing density
matrices of $n$ $d$-level systems, namely $d^{2n} -1 $.

In fact the number of parameters describing LC states will be
less than (\ref{LC_parameters}), since some of the transformations included in
(\ref{general_local}) for a given party will transform pure
states to pure states and so are double-counted in
(\ref{LC_parameters}).  However for the purposes of this letter,
we only require an upper bound and rate of growth with $n$ and $d$
of the number of parameters describing LC states and so
(\ref{LC_parameters}) is adequate.

The above calculations show that for $n>2$, not all density
matrices can be produced by local contamination of pure states
(i.e. not all states are LC states). Specific examples
are given below.  We also note that in the limit of
large $n$ we may say that the dimension of LC states is
essentially the same as that of pure states, at least when
compared to the dimension of the space of all states.  Put another
way, for large $n$,  the space of LC states is exponentially
smaller than the space of all mixed states.

However it is worth noting that for $n$ equal to two, the rate of
growth of the number of parameters describing local contamination
is $d^4$, as is the rate of growth of the number of parameters
describing density matrices. It would be interesting to know
whether every density of two $d$-level systems can arise as the
result of local contamination of pure states. Note that, in order
for it to be possible to reach an arbitrary target density matrix
by local contamination of a pure state, it is necessary that the
number of parameters describing pure states plus contamination be
greater than or equal to that describing density matrices. However
even if parameter counting allows it, this does not guarantee that
it is indeed possible to reach an arbitrary target density matrix;
a more refined analysis is required to determine this.

We now turn to the situation where the parties actively try to
produce a target density matrix; in other words we allow
measurements and classical communication. Our arguments will not
rely on the counting of parameters which is much more subtle here.
To see that there is an issue, consider the simple case of two
$d$-level systems. We can easily show that using classical
communication and local operations, Alice and Bob can produce an
arbitrary density matrix. For let us write the target density
matrix as a mixture of (typically entangled) pure states
$|\psi_\mu\rangle$:
\begin{eqnarray}
\rho = \sum_\mu p_\mu |\psi_\mu\rangle\langle \psi_\mu|.
\label{rho}
\end{eqnarray}
This state may be produced by the following protocol. Alice and
Bob start with a maximally entangled state; we will use this pure state
as the precursor for all target mixed states. We then simply note that
any of the pure states $|\psi_\mu\rangle$ can be produced with
probability one from the maximally entangled state by coordinated
actions by Alice and Bob\cite{Nielsen}. Thus to produce $\rho$, Alice uses a
 random variable to produce an outcome $\mu$ with probability $p_\mu$; she communicates
 the value she receives to Bob; when they
 get the value $\mu$ they
 then perform the protocol to produce  $|\psi_\mu\rangle$ with probability
 one from the maximally mixed state.  Thus overall they have
 produced the required density matrix  $\rho$.  An optimised
 protocol, in the sense of using less entanglement, has been given
 recently by Vidal \cite{Vidal}.

Thus with classical communication Alice and Bob can produce any
density matrix of two $d$-level systems (i.e. every two party
state is LCCC). However as we now show, for more parties, in
general, even with classical communication, not every density
matrix arises as the local contamination of a pure state. For
consider the following state of three qubits
\begin{eqnarray}
Z =p\left| {\rm W}\right\rangle \langle {\rm W}|+(1-p)\left| {\rm GHZ}%
\right\rangle \langle {\rm GHZ}|  \label{Z}
\end{eqnarray}
where $|{\rm W}\rangle = \frac{1}{\sqrt{3}}(\left| 001\right\rangle +\left|
010\right\rangle +\left| 100\right\rangle )$ \cite{Dur00}, and  $|{\rm GHZ%
}\rangle =\frac{1}{\sqrt{2}}(\left| 000\right\rangle +\left|
111\right\rangle )$. Purifications of this state have the form
\begin{eqnarray}
\left| \Psi \right\rangle=\sqrt{p}\left| {\rm W}\right\rangle \left|
f_1\right\rangle +\sqrt{1-p}\left| {\rm GHZ}\right\rangle \left|
f_2\right\rangle,
\end{eqnarray}
where $\left| f_1\right\rangle $ and $\left| f_2\right\rangle $
are orthonormal states of the three ancillas (at this stage we
have made no restrictions about these states; they might be
entangled states of the three local ancilla Hilbert spaces). If it
were possible to create the mixture by local degradation of a pure
state then there must be a pure state $\left| \Phi \right\rangle
$, such that adding local ancillas and evolving with local
unitaries achieves the following transformation
\begin{equation}
\left| \Phi \right\rangle \left| 0\right\rangle \longrightarrow \sqrt{p}%
\left| {\rm W}\right\rangle \left| f_1\right\rangle +\sqrt{1-p}\left| {\rm %
GHZ}\right\rangle \left|f_2\right\rangle.
\end{equation}
Since $|f_1\rangle$ and $|f_2\rangle$ are orthogonal, these two states may
be distinguished with certainty using only local operations and classical
communication \cite{Walgate00}. Thus doing the measurement which distinguishes
$|f_1\rangle$ from $|f_2\rangle$ collapses the state onto $\left| {\rm W}\right\rangle $ or $\left|
{\rm GHZ}\right\rangle$
thus giving some non-zero
probability of creating either $\left| {\rm W}\right\rangle $ or $\left|
{\rm GHZ}\right\rangle $ by LOCC starting with $\left| \Phi \right\rangle $.
However, it has been shown that the three-party states which may be
converted with some probability into a $|{\rm GHZ}\rangle$ state by LOCC,
and those which may be converted into a $|{\rm W}\rangle$ state form two
disjoint classes \cite{Dur00}. Therefore we have a contradiction and so it
is impossible to make the mixture $\rho $ in this way.

We note that our arguments apply equally well to any state of the form (\ref
{Z}) where instead of $\left| {\rm W}\right\rangle $ we have any state in
the $\left| {\rm W}\right\rangle $ class and instead of $\left| {\rm GHZ}%
\right\rangle $ we have any state in the $\left| {\rm
GHZ}\right\rangle $ class. In general, for larger numbers of
particles, there may be a number of non-trivial inequivalent
classes of entangled pure states, and we can produce non-LCCC states
in a similar way. However the above argument will
generalise only for mixtures of two different inequivalent pure
states. This is because it has only been shown that two orthogonal
states can be locally distinguished with certainty. Locally
distinguishing more than two states may require the use of more
copies \cite{Walgate00}.

The details of what happens in the case of more parties and higher spins
remain to be worked out, although the general messages should be clear. In
the case of no classical communication, typical density matrices cannot be
produced by local contamination of pure states. In the case where we allow
classical communication, it is certainly the case that any mixture of two
pure states falling into disjoint classes (as with $\left| {\rm W}%
\right\rangle $ and $\left| {\rm GHZ}\right\rangle $ in the case
of three qubits) will give rise to a ``complex''\ density matrix
(ie. one which cannot be formed by local contamination of a pure
state).  This is a particular method for constructing non-LCCC
density matrices; we suspect that there are many other interesting
classes of non-LCCC states.

The main goal of this letter was to raise the question of how
density matrices can arise.  Once the question has been raised, many
other interesting issues suggest themselves.   A first obvious one
is to find a method for characterising whether a given mixed state
is LC or LCCC.  More generally we would like to understand how the
space of all states decomposes into classes of states which are accessible
from each other by local degrading (with or without classical communication).
We would also like to characterise these classes by
their entanglement properties and assess their physical
implications. Finally, the fact that noise only takes pure states
into a limited range of mixed states, rather than to the whole
space of mixed states (of exponentially larger dimensionality),
offers a quite new perspective on quantum error correction
\cite{errorcorrection}.

\bigskip \noindent{\large {\bf Acknowledgments}}

We gratefully acknowledge funding from the European Union under
the project EQUIP (contract IST-1999-11063).


\begin{references}

\bibitem{many} The number of recent papers dedicated to the study
of density matrices is very extensive.  A necessarily incomplete
list is: R. F.~Werner, Phys. Rev. A {\bf 40}, 4277 (1989),
L.P.~Hughston, R.~Jozsa and W.K.~Wootters, Phys. Lett. A {\bf
183}, 14 (1993), S.~ Popescu,  Phys. Rev. Lett. {\bf 74}, 2619
(1995), A.~Peres, Phys. Rev. Lett. {\bf 77}, 1413 (1996),
M.~Horodecki, P.~Horodecki and R.~Horodecki, Phys. Lett. A {\bf
223}, 1 (1996), C.H.~ Bennett, G.~Brassard, S.~Popescu,
B.~Schumacher, J.A.~ Smolin and W.K.~Wootters, Phys. Rev. Lett.
{\bf 76}, 722 (1996), N.~Linden, S.~Massar and S.~Popescu, Phys.
Rev. Lett. {\bf  81}, 3279 (1998), W. K.~Wootters, Phys. Rev.
Lett. {\bf 80}, 2245 (1998), M.~Horodecki, P.~Horodecki and
R.~Horodecki. Phys. Rev. Lett. {\bf 80}, 5239 (1998), V.~Vedral
and M. B.~Plenio, Phys. Rev. A {\bf 57}, 1619 (1998), G.~Vidal and
R.~Tarrach, Phys. Rev. A {\bf 59}, 141 (1999), S.~Braunstein,
C.M.~Caves, R.~Jozsa, N.~Linden, R.~Schack and S.~Popescu, Phys.
Rev. Lett. {\bf 83}, 1054 (1999), N.~Linden, S.~Popescu and
A.~Sudbery, Phys. Rev. Lett. {\bf 83}, 243 (1999), C. H.~Bennett,
D. P.~DiVincenzo, T.~Mor, P. W.~Shor, J. A.~Smolin and B.
M.~Terhal, Phys. Rev. Lett. {\bf 82}, 5385 (1999), P.~Horodecki,
M.~Horodecki and R.~Horodecki, Phys. Rev. Lett. {\bf 82}, 1056
(1999). Many references to recent literature may also be found in
M.~Lewenstein, D.~Bruss, J.I.~Cirac, B.~Kraus, M.~Kus,
J.~Samsonowicz, A.~Sanpera, R.~Tarrach, J.  Mod. Opt. {\bf 47},
2841 (2000), M.A.~Nielsen and I.L.~Chuang, {\em Quantum
Computation and Quantum Information}, (CUP, Cambridge, UK, 2000).

\bibitem{footnote} In the case of LCCC, if we allow pure state
precursors of higher dimensionality than the target mixed state,
then any mixed state can be trivially obtained: for example by
arranging for maximally entangled states between all pairs of
parties, producing the required mixed state locally at one particular location,
 and then using teleportation.


\bibitem{Schumacher} B. Schumacher,
Phys. Rev. A, {\bf 54}, 2614 (1996).

\bibitem{Nielsen} M.A. Nielsen, Phys. Rev. Lett. {\bf 83}, 436
(1999).

\bibitem{Vidal} G.~Vidal, Phys. Rev. A {\bf 62} 062315 (2000).

\bibitem{Dur00}  W. D\"ur, G. Vidal and J. I. Cirac, Phys. Rev. A {\bf 62}, 062314 (2000).


\bibitem{Walgate00}  J. Walgate, A. J. Short, L. Hardy and V. Vedral,
quant-ph/0007098 (2000).

\bibitem{errorcorrection} P.W.~Shor, Phys. Rev. A {\bf 52} R2493
(1995), A.M.~Steane, Phys. Rev. Lett. {\bf 77} 793 (1996).

\end{references}
\end{document}